\documentclass[preprint,12pt,5p,times,twocolumn]{elsarticle}
\usepackage{epsfig}
\journal{  }
\begin{document}
\begin{frontmatter}
\title{Localized Surface Plasmon Resonance in SnS:Ag Nano-composite
Films}

\author[doe,khalsa]{Priyal Jain}
\author[khalsa]{P.Arun \corref{cor1}\fnref{fn1}}

\cortext[cor1]{Corresponding author}
\fntext[fn1]{(T) +91 11 29258401 (Email) arunp92@physics.du.ac.in}
\address[doe]{Department of Electronic Science, University of Delhi, 
South Campus, Delhi 110021, INDIA}
\address[khalsa]{Material Science Research Lab, S.G.T.B. Khalsa College,\\
University of Delhi, Delhi - 110 007, INDIA}

\begin{abstract}
Nano-composite films of Tin Sulfide (SnS) and silver (Ag) fabricated by 
thermal evaporation showed two prominent peaks in the visible region of their 
extinction spectra. Theoretical modeling of the extinction spectra suggest 
that these two peaks (${\rm \approx 500~nm}$ and ${\rm \approx 580~nm}$) 
correspond to the longitudinal mode (LM) and transverse mode (TM) surface 
plasmon resonance peaks arising from oblate silver nano-particles. Using grain 
size of silver and SnS obtained from structural and morphological 
characterizations of the samples and dielectric constants as per actuals, 
we have compared the experimental results with those from theory. The study
shows that silver nano-particles efficiently scatters light and can be 
used for developing plasmonic based SnS solar cells with improved
efficiencies. 

\end{abstract}
 
%\vfil \eject

\end{frontmatter}

\section{Introduction}
Thin film photo-voltaic devices made of inorganic semiconductors such as Tin
Sulfide (SnS) and Cadmium Telluride (CdTe) etc. have gained attention in 
recent years due to their lower cost compared to the made from Silicon 
\cite{sns1,cdte,Silicon}. SnS is consider a better option considering they 
are not toxic like
Cadmium compounds \cite{cdte2}. SnS is a semiconducting material that 
has a band-gap ranging between 1.1-2.2 eV \cite{bandgap1,sohila} and 
refractive
index in the range of 1.6-2.2~eV, depending on the fabrication techniques 
\cite{refractive,ssc}. However, the major limitation of inorganic 
semiconductors are their 
light trapping ability thereby which their efficiency as solar cells are 
lowered (compared to Silicon) \cite{cat}. In order, to increase the light 
trapping ability of these semiconductors, researchers are adopting a new 
methods using noble metal nano-particles dispersed in the semiconducting
film \cite{cat,cat2}. 

Noble metal nano-particles like silver (Ag), gold (Au), aluminum (Al) and
copper (Cu) embedded in a dielectric/semiconducting medium exhibit interesting 
optical properties and have gained a lot of attention due to their potential 
applications in plasmonic solar cells \cite{cat}, photonic devices \cite{pd}, 
surface plasmon enhanced sensors \cite{sensor}. Localized Surface Plasmon 
Resonance (LSPR) 
occurs when the electromagnetic wave travels along the metal and dielectric/ 
semiconductor interface \cite{pillai}. LSPR peaks and their positions are 
influenced 
by the size, the shape of the metal nano-particle \cite{kelly}, its 
dielectric properties and those of the local surrounding medium \cite{}. For 
nano-particles of Ag, Au and copper the LSPR peak occurs in the UV-Visible 
region of the spectrum depending on the density of the electrons in the metal 
nano-particle\cite{sensor,density1}. Literature consists of many articles on 
metal nano-particles embedded in different dielectric/semiconducting mediums 
like ${\rm SiO_2}$ \cite{sio2}, Si \cite{pillai} and ZnO \cite{zno} etc. Out 
of all the 
noble metals, silver nano-particles showed the best SPR response
\cite{response}, 
making it very useful for applications. We hence selected Ag to study the 
surface plasmon resonance in Tin Sulfide (SnS). Very few research groups have 
carried out work on Ag-SnS films \cite{ag-sns1,ag-sns2,ag-sns3}. In this 
manuscript, we report the 
preparation of SnS:Ag nano-composite films of varying thicknesses and 
investigate the size-dependent optical properties.

\section{Experimental Details}
Composite thin films of Tin sulfide (SnS) and silver (Ag) were grown by
thermal evaporation with vacuum better than ${\rm 4 \times
10^{-5}}$~Torr. The pellets used as starting material were made by mixing 
SnS powder and Ag nano-powder (average grain size 20~nm). The mass ratio of 
SnS and Ag powder used was 2:1. The 99.99\% purity SnS powder was provided by
Himedia (Mumbai) and the Ag nano-powder was obtained from Nanoshel (USA). The 
thickness of the 
films were measured using Veeco's Dektak Surface profiler (150). The 
structural properties of the films were studied using a Bruker D8 
diffractometer at an operating voltage of 40~KV in the 
${\rm \theta-2\theta}$ mode with Cu target giving X-Ray at 
${\rm 1.5418\AA}$ and 
Technai G2 T30U Twin Transmission Electron Microscope (TEM) at an 
accelerating voltage of 200~KV. The chemical composition analysis of the 
films were examined using Kratos Axis Ultra DLD’s X-Ray Photo-electron 
Spectroscopy (XPS) with Al ${\rm k\alpha}$ target. The beam energy used for 
the analysis was 1490~eV and binding energy was referenced with respect to 
Carbon. The surface morphology was analyzed by a Field Emission Scanning 
electron Microscopy (FE-SEM FEI-Quanta 200F). The optical absorption and 
transmittance spectra were recorded using a Systronics UV-VIS Double beam 
spectrophotometer (2202) in the 300-900~nm wavelength range.
%%%%%%%%%%%%%%%%%%%%%%%%%%%%%%%%%%%%%%%%%%%%%%%%%%%%%%%%%%%%%%%%%%%%%%%%%%%%%
\begin{figure}[h!!!]
\begin{center}
\epsfig{file=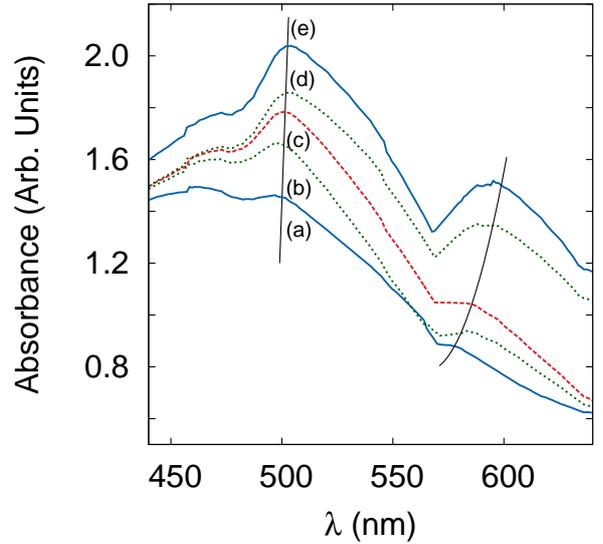, width=3in, angle=-90}
%\hfil
\end{center}
\label{uv}
\caption{\sl Absorption spectra of (a) 270, (b) 480, (c) 600, (d) 650 and (e)
900~nm thick films of SnS:Ag thin composite films.}
\end{figure}
%%%%%%%%%%%%%%%%%%%%%%%%%%%%%%%%%%%%%%%%%%%%%%%%%%%%%%%%%%%%%%%%%%%%%%%%%%%%

\section{Results and Discussion}
\subsection{Experimental Results}
Figure 1 shows the optical absorption spectra of the SnS/Ag composite thin 
films. Unlike the absorption spectra of SnS films \cite{sohila}, these films 
show 
two peaks at ${\rm \approx 500}$ and ${\rm \approx 580~nm}$ in all the films 
without exception. Considering the starting silver
powder had particle size of 20~nm, we believe these peaks are Localized 
Surface Plasmon Resonance (LSPR) peaks resulting from interaction of light 
with electrons at the surface of the silver nano-particles
\cite{singh,awazu}. Considering that the LSPR peak position depends on the 
surrounding media's refractive index along with the shape and size of the 
metal cluster, the existence of two well resolved peaks in fig~1 may either be 
due to non-spherical metallic clusters or because they experience two 
different surrounding media (possibly air and SnS). Another interesting
feature of fig~1 is that both peaks show red-shift, however it is more
pronounced in the ${\rm \approx 580~nm}$ peak.

%%%%%%%%%%%%%%%%%%%%%%%%%%%%%%%%%%%%%%%%%%%%%%%%%%%%%%%%%%%%%%%%%%%%%%%%%%%%%
\begin{figure}[h!]
\begin{center}
\epsfig{file=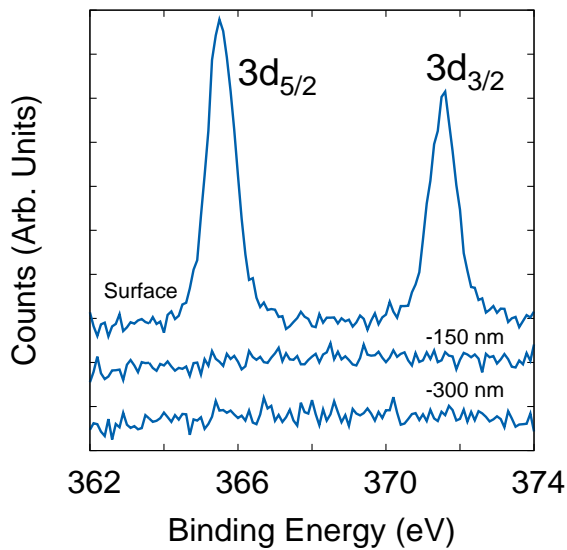, width=3in, angle=-90}
%\hfil
\end{center}
\label{xps}
\caption{\sl XPS scans of Ag 3d peaks at the film surface, 150~nm below the 
surface and 300~nm below the surface. It is clear that silver exists only 
up to 150~nm from the film surface.}
\end{figure}
%%%%%%%%%%%%%%%%%%%%%%%%%%%%%%%%%%%%%%%%%%%%%%%%%%%%%%%%%%%%%%%%%%%%%%%%%%%%
%%%%%%%%%%%%%%%%%%%%%%%%%%%%%%%%%%%%%%%%%%%%%%%%%%%%%%%%%%%%%%%%%%%%%%%%%%%%%
\begin{figure}[h!]
\begin{center}
\epsfig{file=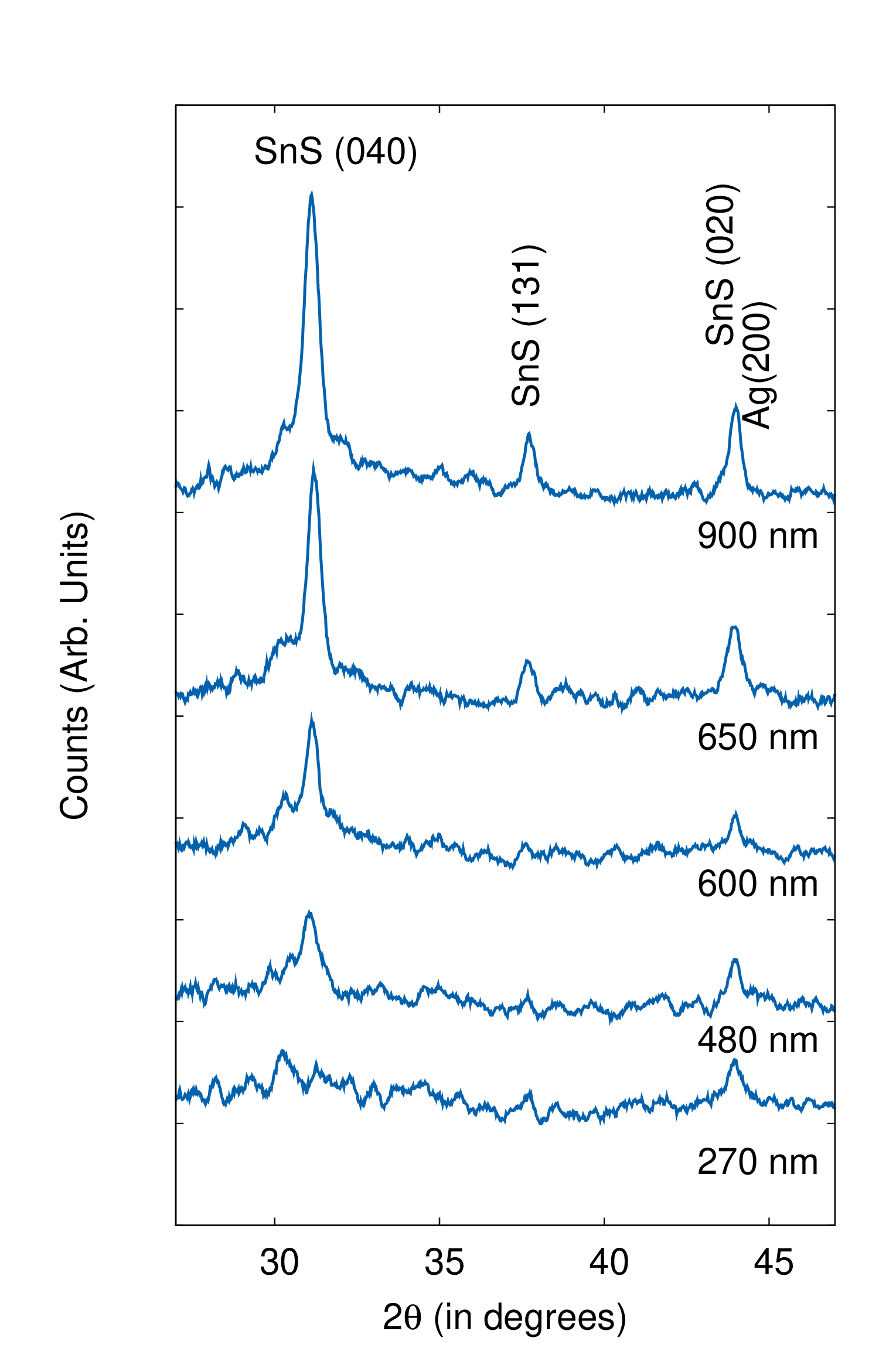, width=5in, angle=-90}
%\hfil
\end{center}
\label{xrd}
\vskip -0.5cm
\caption{\sl 
XRD diffraction patterns for SnS:Ag nano-composite films of various
thicknesses. Prominent planes of SnS and Ag are identifiable and their
Miller indices have been given.}
\end{figure}
%%%%%%%%%%%%%%%%%%%%%%%%%%%%%%%%%%%%%%%%%%%%%%%%%%%%%%%%%%%%%%%%%%%%%%%%%%%%

Before proceeding, we would have to resolve whether silver nano-particles are 
uniformly distributed throughout the thickness of the film or not. We have 
used depth profiling using X-ray Photo-electron 
Spectroscopy (XPS). Fig~2 shows the XPS scan for Ag (Ag 3d peak)
\cite{morales,pilar} 
at the film's surface, at 150 and 300~nm below the surface of a 600~nm thick 
SnS:Ag nano-composite film. Comparing the scans, we find that the XPS peaks 
of silver completely vanishes below 150~nm from the surface of the film. This
clearly implies that the silver nano-particles are only present on and
adjacent to the surface of the films. The deposition of Ag on the surface or 
just below is due to the higher
melting point of Ag (1234~K) compared to (1155~K) that of SnS \cite{CRC,melt}. 
Varying the rate of 
evaporation does give an uniform distribution of silver along the thickness 
of the SnS film. However, since this study is a part of a continuous 
evaluation of SnS and their nano-composite films, we have maintained the 
film fabrication
parameters as in our earlier studies \cite{ssc,jos,tsf}. The existence of 
Ag at the 
surface (in contact with air-SnS interface) and within the SnS surface 
presents a possibility of two LSPR
peaks in the UV-visible absorption spectra which we shall resolve in our
theoretical analysis.
%%%%%%%%%%%%%%%%%%%%%%%%%%%%%%%%%%%%%%%%%%%%%%%%%%%%%%%%%%%%%%%%%%%%%%%%%%%%%
\begin{figure}[h!]
\begin{center}
\epsfig{file=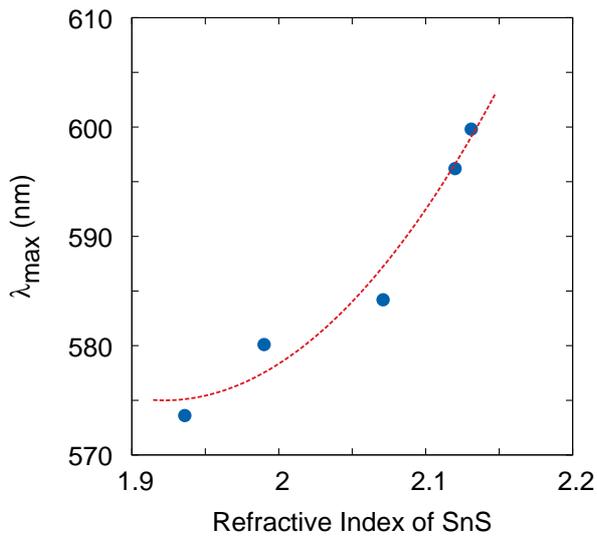, width=3in, angle=-90}
%\hfil
\end{center}
\label{xps}
\caption{\sl Variation of the ${\rm 580~nm}$ peak with increasing refractive
index of the background. Notice that the variation is similar to the trend
line marked in fig~1.}
\end{figure}
%%%%%%%%%%%%%%%%%%%%%%%%%%%%%%%%%%%%%%%%%%%%%%%%%%%%%%%%%%%%%%%%%%%%%%%%%%%%
%%%%%%%%%%%%%%%%%%%%%%%%%%%%%%%%%%%%%%%%%%%%%%%%%%%%%%%%%%%%%%%%%%%%%%%%%%%%%
\begin{figure}[h!]
\begin{center}
\epsfig{file=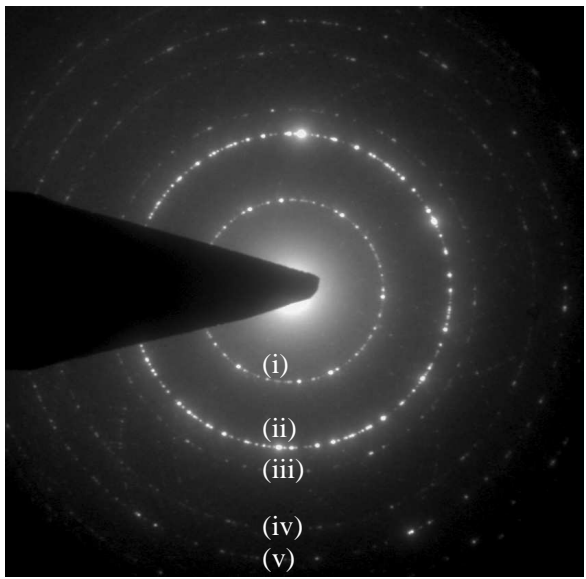, width=3in, angle=-0}
\end{center}
\label{saed}
\caption{\sl Shows the Selected Area Diffraction Pattern of a 600~nm thick film. 
The brighter inner rings (i) and (ii) corresponds to those of Ag diffraction
planes while the remaining rings (iii, iv and v) are those of SnS planes. }
\end{figure}
%%%%%%%%%%%%%%%%%%%%%%%%%%%%%%%%%%%%%%%%%%%%%%%%%%%%%%%%%%%%%%%%%%%%%%%%%%%%

As stated above, the 580~nm absorption peak of fig~1 shows a strong
red-shift which increases with increasing film thickness. An 
interesting aspect about LSPR peaks is that its position strongly depends on 
the refractive index of the surrounding media. An increasing refractive index 
of the surrounding dielectric results in a red-shift \cite{shift}. Hence, 
clearly, 
we expect that the SnS film's refractive index increases with increasing film
thickness. To investigate the possible cause of increase in SnS refractive 
index with film thickness, we investigate the structural properties of 
samples. Fig~3 shows the diffraction patterns of various thickness SnS:Ag
nano-composite films. Without exception all the films showed
polycrystallinity with SnS having orthorhombic structure with lattice
parameters matching those given in ASTM~Card No 83-1758 while Ag matched those given
in ASTM~Card No 03-0931. The grain size of SnS and Ag nano-particles were 
evaluated using the Scherrers formula \cite{cullity}. We shall be commenting 
on the grain size of silver nano-particles below, however the SnS
grains showed a linear increase in grain size with increasing film
thickness. In earlier studies \cite{ssc,jos,tsf} we conducted on thin SnS 
films (without Ag) grown under identical conditions, we had reported the 
optical properties of SnS films as a function of grain size. We had reported 
that the SnS film's refractive index increased with increasing grain size, 
which in turn increased with film thickness (fig~6 of 
ref~\cite{ssc}). We find that the SnS grain size (12-19~nm) and its
variation with film thickness in these nano-composite films matched (within
experimental error) with previous studies. As stated initially, we believe 
that the increasing refractive index with increasing thickness of 
SnS background contributes to the observed red-shift in the 580~nm LSPR peak
of fig~1. As explained above, we use experimentally evaluated refractive
index of SnS films as a function of film thickness and plot position 
of the ${\rm \approx 580~nm}$ SPR peak (${\rm \lambda_{max}}$) with respect
to it (fig~4). Notice that the variation is similar to that shown in fig~1. 
The experimentally evaluated values of refractive index would hence be
utilized as input parameters for the Gans model calculations we discuss in the
next section. Also, we notice only a small variation in refractive index
with SnS grain size for ${\rm \lambda \approx 500~nm}$ explaining the
insignificant or minor red-shift seen in the ${\rm \approx 500~nm}$ peak.

%%%%%%%%%%%%%%%%%%%%%%%%%%%%%%%%%%%%%%%%%%%%%%%%%%%%%%%%%%%%%%%%%%%%%%%%%%%%%
\begin{figure}[h!]
\begin{center}
\epsfig{file=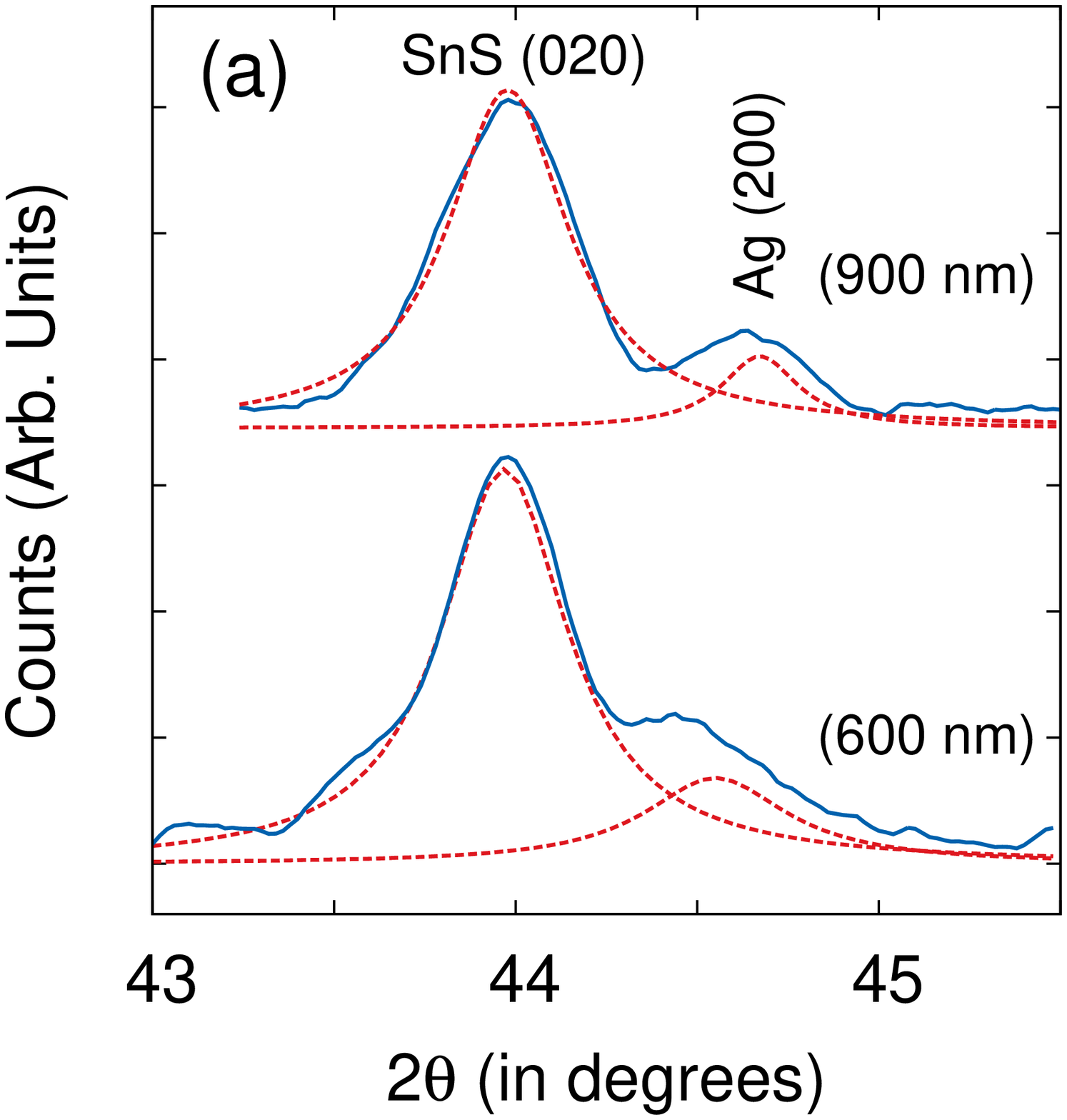, width=2.25in, angle=-90}
\hfil
\epsfig{file=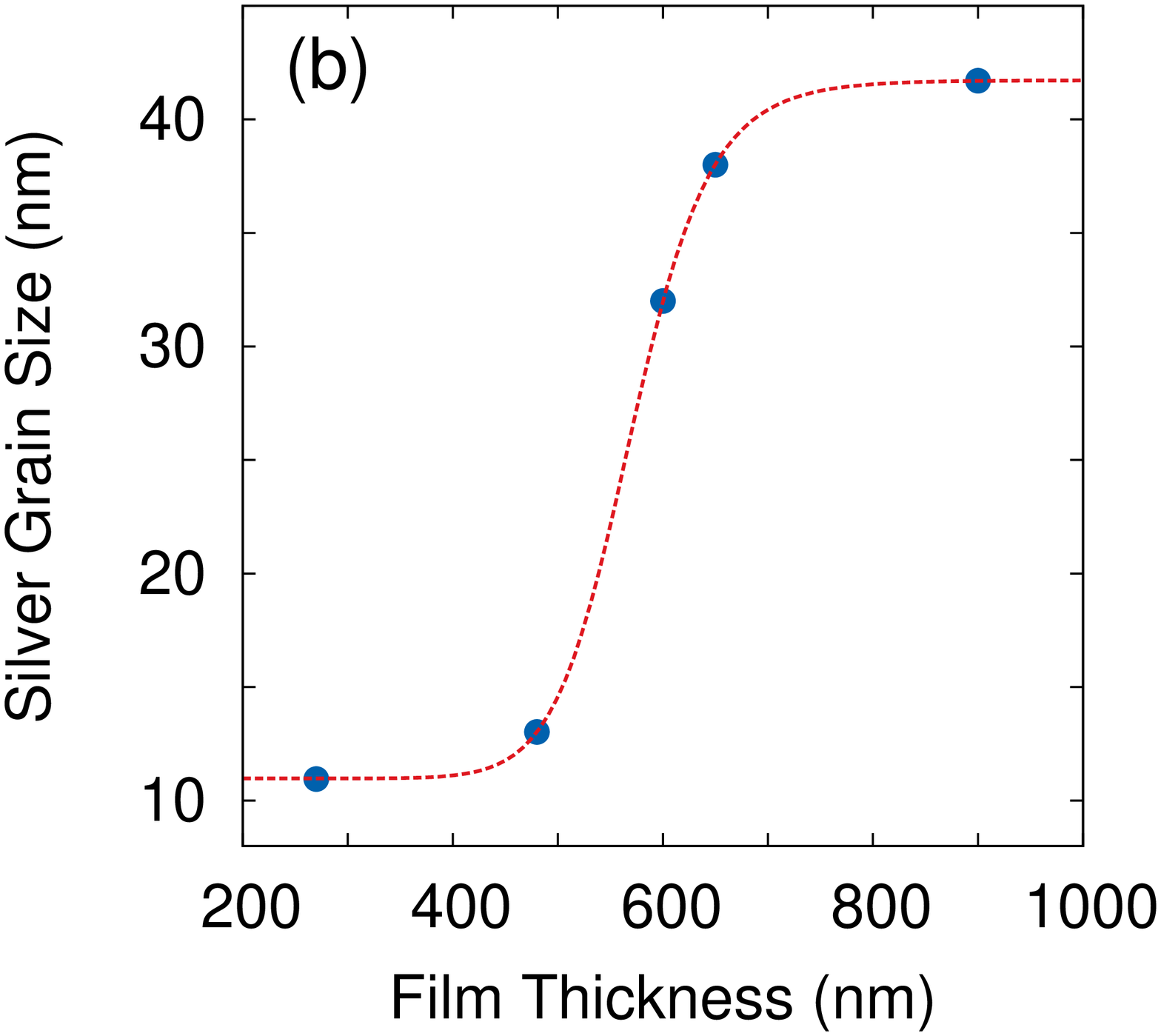, width=2.25in, angle=-90}
\end{center}
\label{sem}
\caption{\sl Silver grain size were calculated using Scherrers formula after
deconvoluting the overlapping yet resolvable XRD peak around ${\rm 44^o}$
(a). The thicker SnS:Ag nano-composite films (greater than 600~nm) have grain
size of the order of 32-42~nm (b).}
\end{figure}
%%%%%%%%%%%%%%%%%%%%%%%%%%%%%%%%%%%%%%%%%%%%%%%%%%%%%%%%%%%%%%%%%%%%%%%%%%%%

%%%%%%%%%%%%%%%%%%%%%%%%%%%%%%%%%%%%%%%%%%%%%%%%%%%%%%%%%%%%%%%%%%%%%%%%%%%%%
\begin{figure}[h!!!]
\begin{center}
\epsfig{file=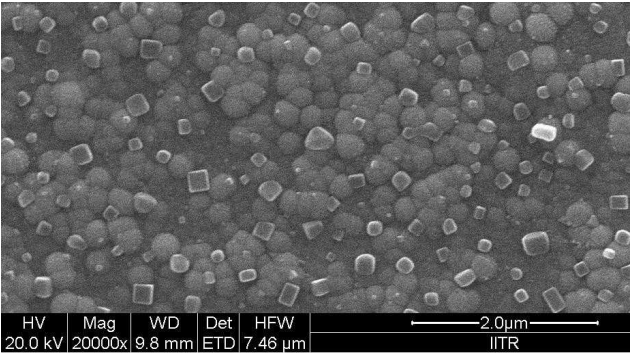, width=3.25in, angle=-0}
\end{center}
\label{sem}
\caption{\sl SEM micrograph showing surface morphology of a 650~nm thick
SnS:Ag nano-composite film. The spherical grains in the background are that
of SnS while the sharp cuboid grains are that of silver.}
\end{figure}
%%%%%%%%%%%%%%%%%%%%%%%%%%%%%%%%%%%%%%%%%%%%%%%%%%%%%%%%%%%%%%%%%%%%%%%%%%%%

Other then the XPS results, the lone XRD peak of Ag also confirms the 
existence of metallic clusters of silver in the nano-composite films.
Further proof of metallic Ag clusters in our samples come from the electron 
diffraction rings (Selected Area Electron Diffraction, SAED 
image) obtained using the Transmission Election Microscope. SAED pattern
(fig~5) 
shows two sharp rings corresponding to two prominent peaks from Ag. The two 
Ag rings marked (i) and (ii) in fig~5 correspond to the (200) and
(220) planes. The remaining three rings of fig~5, marked (iii), (iv) and
(v), correspond to the (040), (131) and (200) diffraction planes of SnS. 
An investigation of the diffraction patterns at ${\rm 2\theta \sim 39^o}$ and 
${\rm \sim 44^o}$, of fig~3, shows them to be resolvable peaks of SnS and Ag 
(fig~6a). The peaks were de-convoluted and the average grain size of silver 
nano-particles were evaluated using the Scherrer's formula. Fig~6b shows the 
variation in the silver nano-particle grain size with film thickness. 
Notice that the average size of the Ag nano-particle is
around 12~nm for films with thickness less than 600~nm while they are in the
range of 32-42~nm for higher thickness films. As the size of the metal
cluster increases there is an increase in radiative damping that leads to
greater scattering cross-sections and hence larger extinction \cite{cat}.

Fig~7 exhibits a SEM micrograph of a 650~nm thick SnS:Ag nano-composite 
film. As viewed from the surface, micrograph shows that the film consists of 
randomly spaced cuboid silver
nano-particles \cite{wiley} dispersed among nearly spherical SnS 
nano-particles of SnS. 
The chemical composition of the grains were confirmed using the 
EDAX (Energy Dispersive Analysis of X-ray) attachment with the scanning
electron microscope. 

\subsection{Theoretical Consideration}

The extinction spectra of fig~1 can be theoretically generated using various
computational models, like Mie Model \cite{mie1}, Gans Model
\cite{gans1,gans2} and Discrete Dipole Approximation (DDA) Model 
\cite{dda1,dda2}. Mie theory was developed 
to study scattering due to spherical particle \cite{mie1}. The model was 
successfully extended for studying Localized Surface Plasmon Resonance due to 
spherical metal nano-particles dispersed in semiconducting or insulating
media. However, the model needed corrections to include non-spherical shaped
metal clusters. The modifications proposed by Gans \cite{gans1,gans2}. 
Considering that our SEM morphology images suggest our 
silver metal clusters to have near square cross-section on the surface 
(${\rm a \approx b}$) 
and XPS results suggest the grains to occur only on the film surface, we 
expect ${\rm a \approx b > c}$ i.e. the silver nano-particles are oblate 
shaped with size less than 50~nm. We have, hence used Gans model to 
theoretically analyze our experimental results. 

Gans model gives the extinction cross-section/ coefficient which is the sum 
of light intensity scattered and absorbed by a metal nano-particle whose 
shape is either oblate or prolate. The scattering and absorption 
cross-sections are calculated using \cite{cat}
\begin{eqnarray}
\sigma_{scat} =\left({ 16\pi^3 \over
6\lambda^4}\right)Re(\alpha)^2\label{second}
\end{eqnarray}
and 
\begin{eqnarray}
\sigma_{abs} = \left({2\pi \over \lambda}\right) Im(\alpha)\label{third}
\end{eqnarray}
where,
\begin{eqnarray}
\alpha = \left({\epsilon_o V\over L}\right) \left[{\epsilon-\epsilon_m 
\over \epsilon + \left({1-L\over L}\right) \epsilon_m}\right]\label{first}
\end{eqnarray}
where, `L' is the depolarization factor that depends on the shape of the 
scattering/ absorbing metal nano-particle, `V' is the volume of the 
nano-particle, `${\rm \epsilon}$' is the complex dielectric constant of the 
metal and `${\rm \epsilon_m}$' is that of the surrounding medium. For 
non-sphere clusters, the depolarization factor `L' takes different values 
along the three axes. The values completely depend on the shape of the metal 
nano-particle which is mathematically defined by the aspect ratio 
(ratio of minor to major axis of the spheroid). For an oblate nano-particle 
(${\rm a \approx b>c}$) it is given as
\begin{eqnarray}
L_z &=& \left({1 + e^2\over e^3}\right)[e-\tan^{-1}(e)]\nonumber\\
L_x &=& L_y = \left({1-L_z \over 2}\right)\nonumber
\end{eqnarray}
Where `e' is related to the aspect ratio 
\begin{eqnarray}
e = \sqrt{1-\left({c \over a}\right)^2}\nonumber
\end{eqnarray}
Eqn~\ref{first} is now written as
\begin{eqnarray}
\alpha = \left({\epsilon_o V\over 3L}\right) \sum_{i=1}^{3} 
\left[{\epsilon-\epsilon_m \over \epsilon + \left({1-L_i\over L_i}\right) 
\epsilon_m}\right]\label{sumed}
\end{eqnarray}
When an E-M radiation is incident on a spheroid nano-cluster, electrons are
set into oscillations along and perpendicular to its symmetry axis,
resulting in two plasmon resonance peaks. Each peak is associated with the
oscillation direction with respect to the symmetry axis and are named as the 
longitudinal mode (LM) and the transverse mode (TM) \cite{noguez}. In the case 
of oblate nano-clusters the TM excitations appear at higher wavelengths
compared to that of the LM excitations \cite{noguez}. Expectedly, the 
separation between the two modes depend on how much deviation the spheroid 
has from an
ideal sphere, or in other words depends on the cluster's aspect ratio and the 
refractive index of the surrounding medium \cite{kelly,noguez}. 
%%%%%%%%%%%%%%%%%%%%%%%%%%%%%%%%%%%%%%%%%%%%%%%%%%%%%%%%%%%%%%%%%%%%%%%%%%%%%
\begin{figure}[t!!!]
\begin{center}
\epsfig{file=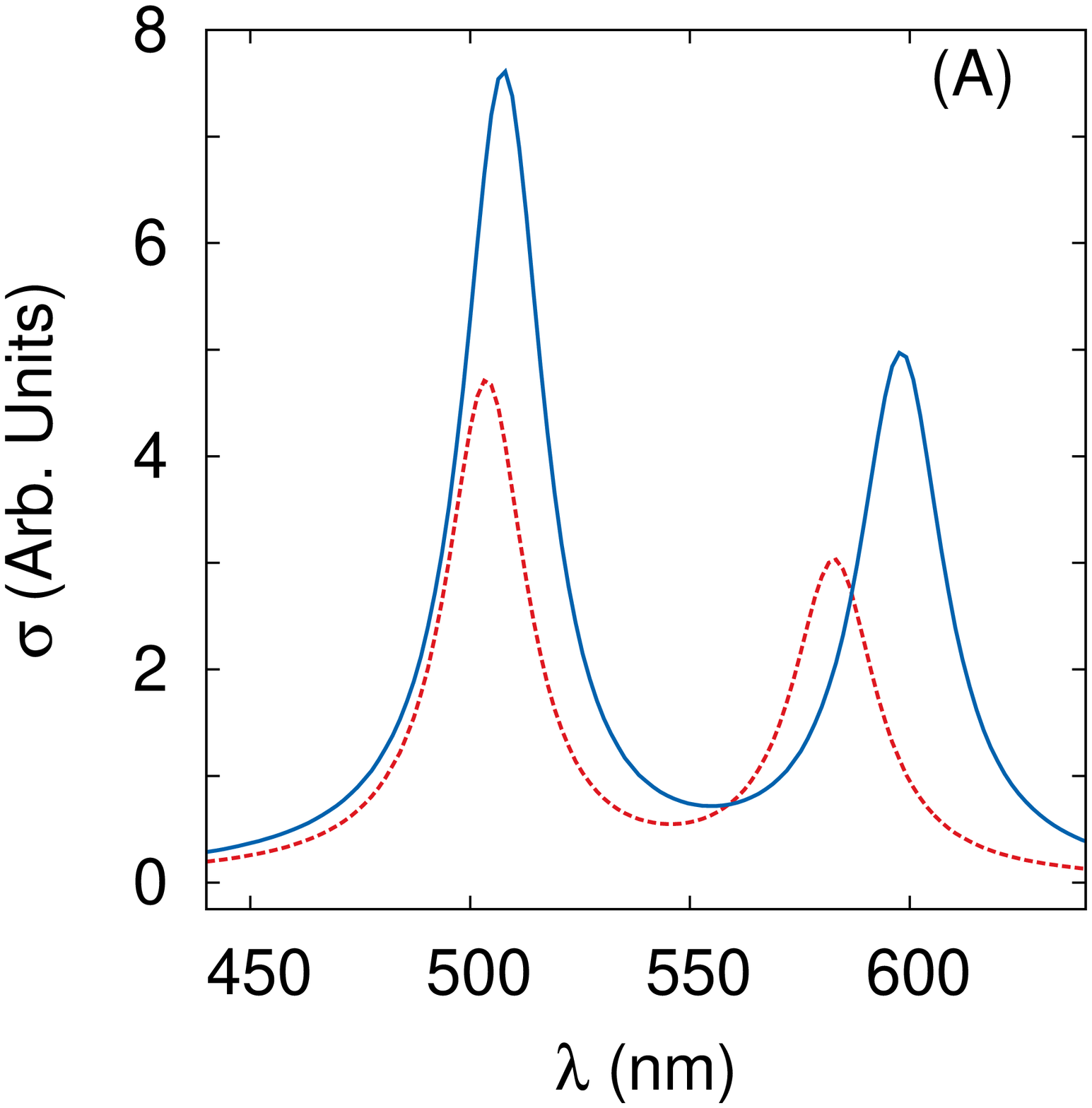, width=2.5in, angle=-90}
\hfil
\epsfig{file=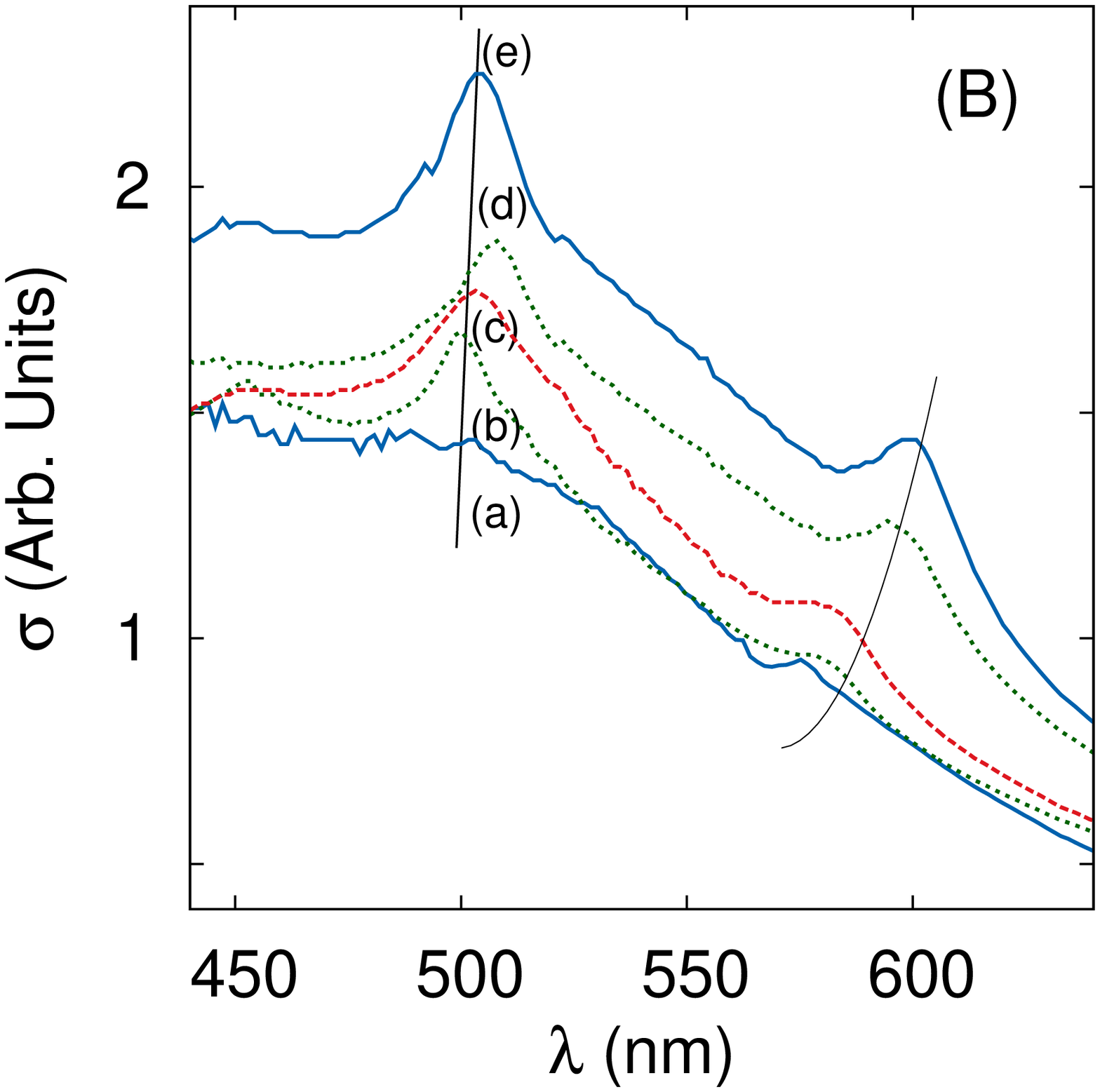, width=2.5in, angle=-90}
\end{center}
\label{scatterGS}
\caption{\sl (A) Shows the results of theoretical calculations of extinction
coefficients for Ag nano-clusters in SnS medium in 650 and 900~nm films 
respectively. (B) shows the combined result of absorption by SnS films
(experimental data of SnS films used) and extinction by silver nano-clusters
(as theoretically calculated).}
\end{figure}
%%%%%%%%%%%%%%%%%%%%%%%%%%%%%%%%%%%%%%%%%%%%%%%%%%%%%%%%%%%%%%%%%%%%%%%%%%%%
Our calculations for extinction coefficients were done using the complex
dielectric constant for silver listed as a function of wavelength by Palik et 
al \cite{palik}. We used the data from fig~6 for estimating the grain size of 
silver. Before calculating case of silver dispersed in SnS, simulation was
done for silver in air. Our calculations agreed with literature \cite{zno} and
found LSPR peak appear at 350~nm. Thus, we can confidently rule out
silver-air interface playing a role in this study. We hence proceed
reporting effect of Ag-SnS interface in this study. The 
results of Priyal et al \cite{jos,ssc} were used for the SnS background's 
dielectric constant. The studies showed SnS film's refractive index follow a
trend similar to the Cauchy's dispersion relation \cite{elkorashky} with the 
wavelength, where the coefficients strongly depend on SnS grain size. Fig~8 
shows the 
extinction spectra generated theoretically. The aspect ratio was linearly 
varied
from 0.64 to 0.60 with increasing film thickness. The values were selected 
so that peak positions obtained theoretically
matched with the experimentally obtained results. Fig~8A shows select
calculated extinction coefficient spectra. These spectra are of course quite
dissimilar from those shown in fig~1 since fig~1 also contains the
absorption spectra of the background SnS film. To recreate fig~1 we have
super-imposed data of fig~8A on absorption spectra of SnS films of
corresponding thicknesses with same/similar grain size used in our previous
studies \cite{jos}. We may hence 
conclude that the peaks of fig~1 at ${\rm \sim 500}$ and 
${\rm \sim 580~nm}$ correspond to the longitudinal and transverse modes 
respectively. Fig~8B gives thus theoretically simulated spectra and we
find them in fairly good agreement with the experimentally obtained data of
fig~1. Fig~8B also includes the two trend lines of fig~1 to show how well
our simulations agree with experimental data.

The above results clearly establishes the formation of LSPR peaks in SnS:Ag
nano-composite films. It is of immense interest that the well established
theory easily quantifies the experimental observations. As stated, SnS films 
have attracted interest
in recent years for its application as photovoltaics \cite{noguchi,ghosh}. 
While their
properties of being non-toxic and films with good absorption and high
band-gap make them suitable candidates, attempts of fabricating
photovoltaics show that they do not match the efficiency of silicon
\cite{noguchi,david}.
A basic issue with thin film solar cell devices is that the absorbing layer 
being thin does not absorb sufficient amount of light for charge carrier 
generation and hence leads to low efficiency. Hence, research has been 
directed towards plasmonic solar cells of SnS. One expects that the metal 
nano-particles in the medium would at resonant frequency either absorb the
incident light or scatter it within the medium \cite{cat,jain}. If 
scattering can be made the dominant extinction mechanism, the efficiency of 
the device can be enhanced \cite{cat}. Using the above made calculations 
(namely 
using eqn~\ref{second} and \ref{third}), we have calculated the scattering
efficiency (${\rm \eta}$) using
\begin{eqnarray}
\eta (\%)= {\sigma_{scat}\over \sigma_{scat} + \sigma_{abs}}\times 100\nonumber
\end{eqnarray}
%%%%%%%%%%%%%%%%%%%%%%%%%%%%%%%%%%%%%%%%%%%%%%%%%%%%%%%%%%%%%%%%%%%%%%%%%%%%%
\begin{figure}[t!!!]
\begin{center}
\epsfig{file=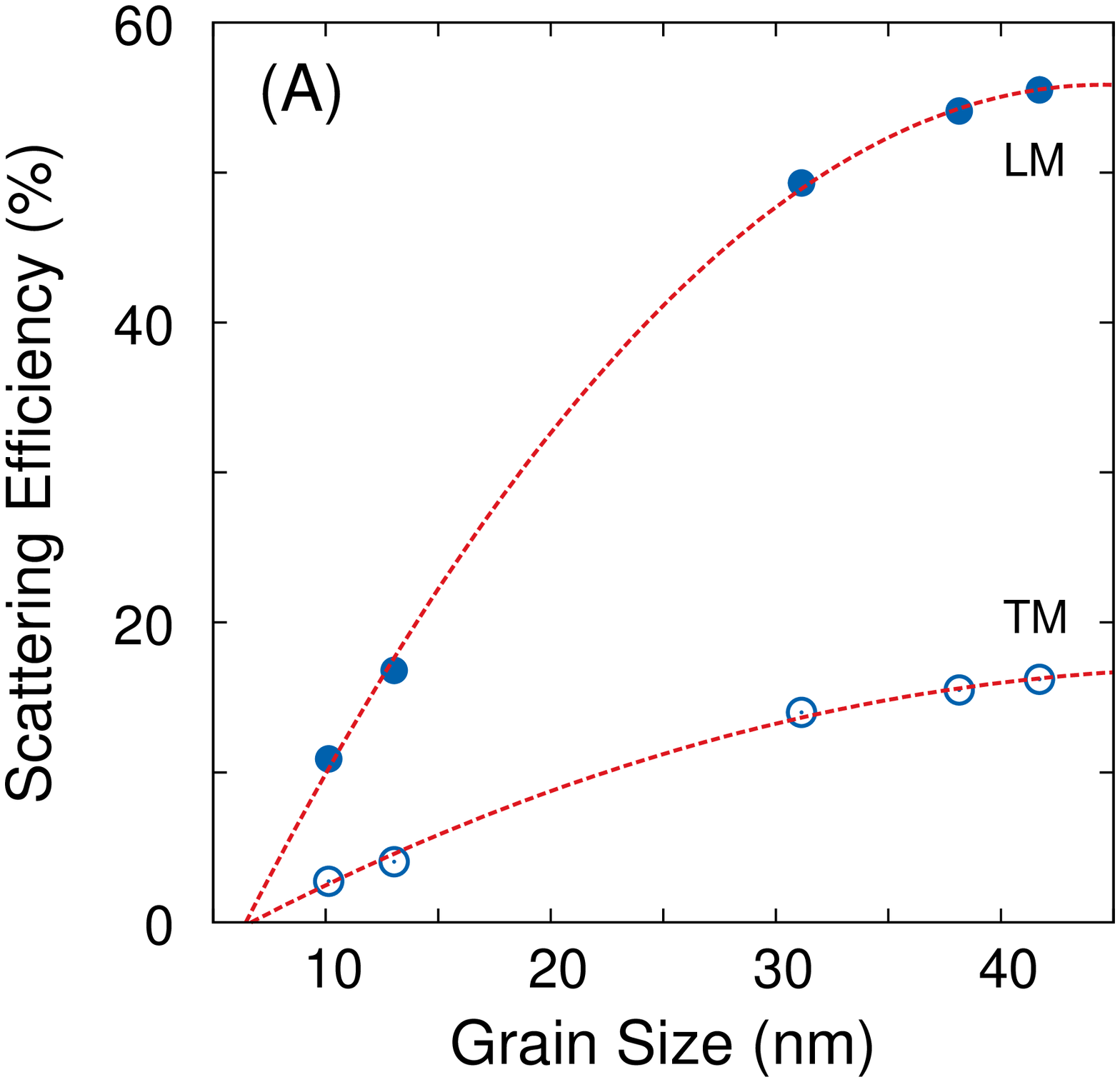, width=2.5in, angle=-90}
\hfil
\epsfig{file=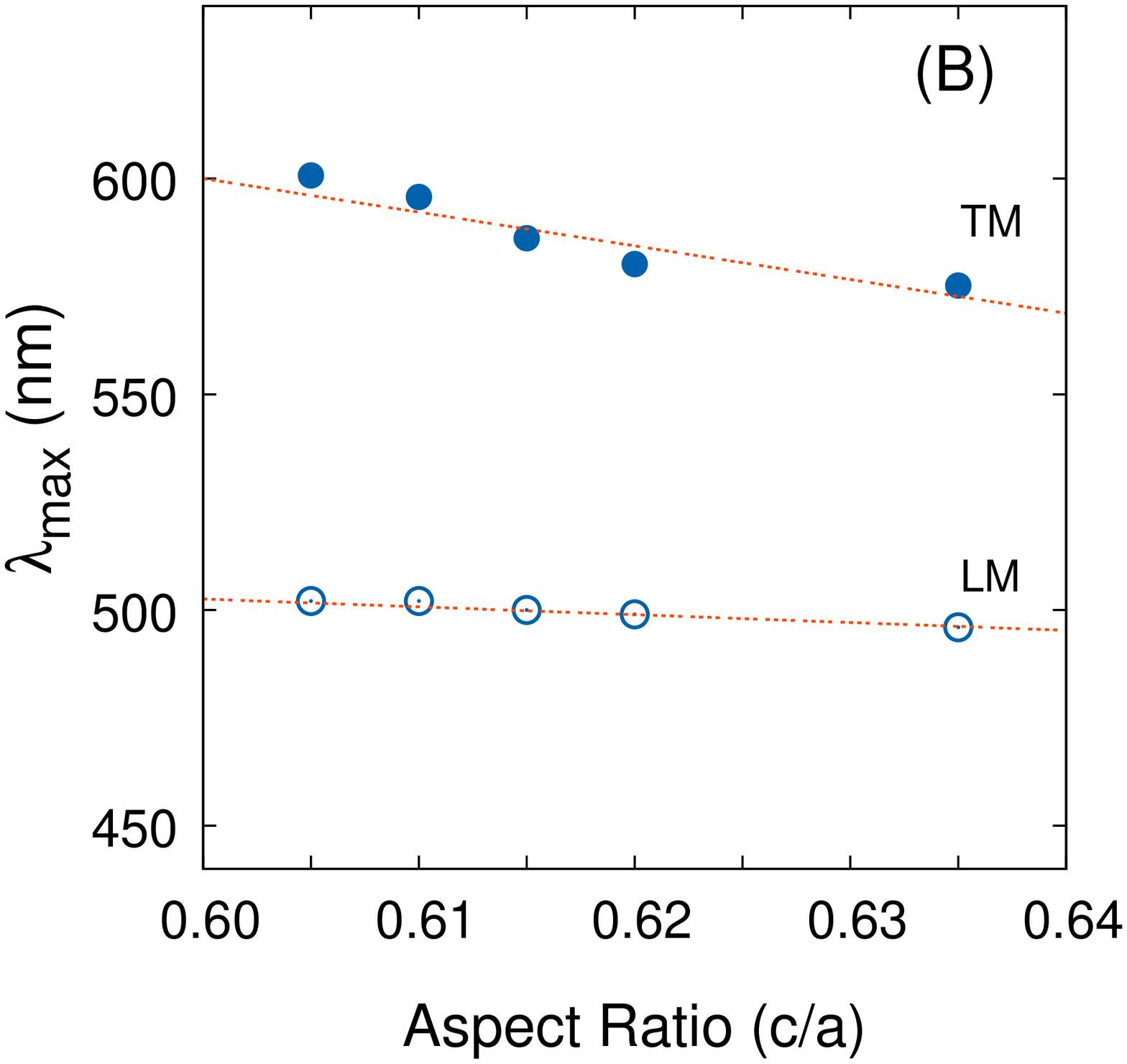, width=2.5in, angle=-90}
\end{center}
\label{scatterGS2}
\caption{\sl (A) shows the variation in scattering efficiency at wavelengths
corresponding to peaks of TM and LM with grain size. Variation of (B) and
fig~4 clearly shows the grain dependence of (A) comes from the SnS
background's refractive index and silver grain's aspect ratio (in case of
TM).}
\end{figure}
%%%%%%%%%%%%%%%%%%%%%%%%%%%%%%%%%%%%%%%%%%%%%%%%%%%%%%%%%%%%%%%%%%%%%%%%%%%%
Fig~9A shows scattering efficiency of the LM and TM peaks. The scattering 
efficiencies vary as a (quadratic) function of the grain size and are 
marked different for the LM and TM modes. The scattering contribution is
more in the LM peak as compared to that of the TM peak. Fig~9A is misleading
since grain size only contributes to extinction's intensity via the volume
term of eqn~\ref{first}. Size indirectly influences the scattering efficiency 
through the real component of silver's 
dielectric constant (${\rm \epsilon}$) and SnS background's dielectric
constant (${\rm \epsilon_m}$). Both these factors vary with respective grain 
size (silver
and SnS) which are in turn dependent on the film thickness. As the film 
thickness increase, the grain size of silver and SnS increases. This results
in increased difference in silver's and SnS's dielectric constant, resulting
in stronger scattering from the metal nano-particles (also understood from
eqn~\ref{sumed}). This is possibly due to 
the fact that the effective size of the metal nano-particle increases as the 
refractive index of SnS increases leading to larger scattering. The result
is promising considering that we now can predict and material manipulate
plasmonic layer to obtain high efficiency solar devices.
Another aspect of interest for device fabrication would 
be to bring the two resolved peaks close by. This would result in an
absorption band leading to a range of wavelengths being captured and
scattered within the medium encourage easy electron-hole pair generation. It
is the non-spherical shape of the metal nano-clusters (aspect ratio ${\rm
\neq 1}$) that gives rise to the two peaks. This is best understood by
investigating the peak positions (${\rm \lambda_{max}}$) as a function of
aspect ratio (fig~9B). Both, the LM and TM peaks show blue shifts
with increasing aspect ratio (c/a), this trend is in agreement to those
reported in literature \cite{noguez}. The dotted lines indicate the trends and
suggest that they would tend to converge around 0.75, for which we would
have the TM and LM peaks close by resulting in a board absorption band.
 
In conclusion, we can relate all of the above results. For an efficient
photo-layer of a solar device, we ideally require SnS:Ag nano-composite films 
with 40~nm sized silver metal nano-clusters which are oblate shaped with 
aspect ratio of 0.75 and large sized SnS grains. This ensures a large 
difference in SnS and silver refractive
indices that lead to enhanced scattering of light within the films and the 
aspect ratio ensure
unresolved TM and LM peaks giving broad absorption band.

\section*{Conclusion}
Nano-composite films of SnS and Ag were grown on glass substrates at room
temperature. The films showed two prominent peaks at ${\rm \rm \approx
500~nm}$ and ${\rm \approx 580~nm}$ in their UV-visible spectra. These peaks 
showed a red-shift as the film thickness increased. Structural studies reveal 
increase in Ag and SnS nano-particle size with increasing film thickness. 
We were able to generate the observed extinction spectra using Gans model. 
The results of our calculations suggest that the ${\rm \approx 500~nm}$ and 
${\rm \approx 580~nm}$ corresponds to the longitudinal and transverse mode 
LSPR peaks from oblate silver 
nano-particles. The theoretical and experimental results also explains the 
observed red shift in the SPR peak at ${\rm \approx 580~nm}$ due to the 
increasing refractive index of the background SnS. Theoretical results 
also suggests that as the aspect ratio of Ag nano-particles decreases with 
increasing film thickness, the two resonant peaks move away from each other. 
The calculations also explain the increase in scattering efficiency with 
increasing Ag nano-particle size as resulting from enhanced difference
between the metal and its background's dielectric constant. 
Both theoretical and experimental results suggest that one can tailor the SPR 
peak positions and scattering efficiency of such nano-composites. SnS:Ag
nano-composite films hence present itself as an efficient photo-layer in
solar voltaics. Based on our results we suggest SnS:Ag nano-composite 
films with oblate shaped (aspect ratio ${\rm \approx 0.75}$) Ag nano-particles
of size around 40~nm which embedded in large grain sized SnS background
would prove ideal for plasmonic devices.

\section*{Acknowledgement}
Authors are thankful to the University Grants Commission (UGC, India) for 
funding this work under their Major Research Project Scheme (F.No.
39-531/2010SR).

\bibliographystyle{model1a-num-names}
%\bibilography{00}

\end{document}